\documentstyle[preprint,aps]{revtex}

\begin{document}
\draft
\preprint{SNUTP-94-104}
\title{Suppression of dilepton production in hot hadronic matter}
\bigskip
\author{Chungsik Song$^1$\footnote{e-mail: song@comp.tamu.edu},
 Su Houng Lee $^2$\footnote{e-mail: suhoung@phya.yonsei.ac.kr},
 and Che Ming Ko  $^1$\footnote{e-mail: ko@comp.tamu.edu}}
\address{\it $^1$Cyclotron Institute, Texas A\&M University, College Station,
 TX 77843, USA \\
 $^2$Department of Physics, Yonsei University, Seoul 120-749, Korea}
\maketitle
\begin{abstract}

The pion electromagnetic form factor at finite temperature is studied
using an effective chiral lagrangian that includes explicitly
vector mesons.  We find that in the time-like region
around the rho meson resonance it decreases with increasing temperature
and leads to a suppression of
dilepton production from pion-pion annihilation in a hot hadronic matter.
Effects on dilepton production in high energy heavy ion
collisions and its relevance to the phase transition in a hot hadronic matter
are discussed.

\end{abstract}
\vspace{0.5cm}
\pacs{PACS numbers : 25.75+r, 13.40.Fn, 11.30.Rd}


Dileptons from relativistic heavy ion collisions
have continually attracted great interest as once produced they would
escape from the collision region without further interaction and
are thus ideal probes of the hot dense matter formed in the initial stage
of the collision \cite{dilepton}.   In hot dense matter, chiral
symmetry is expected to be partially restored and the deconfinement
transition to
the quark-gluon plasma is also possible if the temperature and density
are sufficiently high. Dileptons can therefore provide
the signatures for these new phases of hadronic matter.

It has been suggested that the masses of vector mesons would change as the
hadronic matter undergoes a phase transition to the chirally symmetric phase
\cite{pisaski}.
If this is the case, one should be able to observe this effect directly
through the shift of vector meson peaks in the dilepton spectra from
heavy ion collisions \cite{hatsuda}.
However, based on PCAC and current algebra it has been shown
that up to $T^2$, where $T$ is the temperature,
there is no change in vector meson masses but
only a mixing between the vector and axial vector correlators \cite{dei}.
This result should
be satisfied by any models that include
the symmetry properties of low energy hadronic physics.
Indeed, results from both QCD sum-rule calculations \cite{FHL,EI94}
and effective chiral lagrangian approaches \cite{slee}
are consistent with this temperature dependence, and
vector meson masses obtained from these models do not change appreciably
unless the temperature of the hadronic matter is very
close to the critical temperature for the phase transition.

On the other hand, the yield of dileptons from hot matter may have
a stronger dependence on temperature.
In hot hadronic matter, the production of
dileptons with invariant masses
near the $\rho$ resonance is dominated by
pion-pion annihilation.  According to vector meson dominance
(VMD) \cite{sak}, two pions in this process
form a rho meson that subsequently converts into a virtual photon.
The dilepton yield depends thus on the
pion electromagnetic form factor,
\begin{equation}
F_\pi(q^2)={{g_{\rho\pi\pi} g_\rho}
\over m_\rho^2-q^2-im_\rho\Gamma_\rho},
\end{equation}
where $g_\rho$ is the photon-$\rho$-meson coupling constant,
$g_{\rho\pi\pi}$ is the pion-$\rho$-meson coupling constant,
and $\Gamma_\rho$ is the neutral $\rho$ meson decay width.
This form factor has been extensively used in calculating
the dilepton emission rate from hadronic matter at finite temperature
\cite{siemens,russ,kapusta,ko,matsui,levai,lichard}.
In these studies, the form factor has been taken to be independent
of temperature. However, recent studies using QCD sum-rules \cite{domi} or
QCD factorization \cite{satz} have suggested that $F_\pi(q^2)$ is likely to
be modified at finite temperature, and this will have effects
on dilepton production in hot hadronic matter.

In the present paper, we shall study the pion electromagnetic form
factor at finite temperature using an effective chiral lagrangian
that includes explicitly the vector mesons and
gives also the correct isospin mixing at finite temperature.
Our study indicates that not only is the photon-$\rho$-meson
coupling modified at finite temperature
but there also exist effects due to vertex corrections and
changes of pion and $\rho$ meson properties in the hot hadronic matter.
Using the temperature-dependent form factor, we shall study its
effect on the dilepton production rate from $\pi-\pi$ annihilation in
hot matter.


Although the chiral perturbation theory \cite{Wein,GL84} has been
successful in describing systematically low energy hadronic
phenomena, it has not been able to account for higher energy processes
related to vector mesons. In the literature,
two methods have been introduced to include vector
mesons and photon fields in the chiral lagrangian;
the massive Yang-Mills approach \cite{MYMA} and the hidden gauge approach
\cite{BKY}.
In the massive Yang-Mills approach, $\rho$ and  $a_1$ mesons are
introduced as external gauge fields of the chiral group
and the photon field is introduced via VMD \cite{MYMA}.
In the hidden gauge approach, vector mesons are introduced
as gauge fields of the hidden local symmetry and the photon is
introduced as an external gauge field \cite{BKY}.
These two methods have been shown to be gauge
equivalent \cite{Yamawaki1,Zahed1} and to
have identical symmetry properties at finite temperature \cite{slee}.

We shall follow the hidden gauge approach by considering the
$[SU(2)_{\rm L} \times SU(2)_{\rm R}]_{\rm global}$ $\times$
$[SU(2)_V]_{\rm local}$ ``linear" sigma model.  It is constructed with two
SU(2)-matrix valued variables $\xi_L(x)$ and $\xi_R(x)$, which transform
as $\xi_{L,R}(x) \rightarrow \xi'_{L,R}(x)=h(x) \xi_{L,R}~ g^\dagger_{L,R}$
under
$h(x)\in[$SU(2)$_V]_{\rm local}$ and $g_{L,R}\in[$SU(2)$_{L,R}]_{\rm global}$.
Introducing the vector meson $V_\mu$  as the gauge field of the local
symmetry and the photon ${\cal B}_\mu$ as an external gauge field of the global
symmetry, we have the following chirally invariant lagrangian,
\begin{eqnarray}
{\cal L}& = & f_\pi^2 {\rm tr} \left[\frac{1}{2i} ( {\cal D}_\mu \xi_L \cdot
         \xi_L^\dagger - {\cal D}_\mu \xi_R \cdot \xi_R^\dagger) \right]^2
 \nonumber \\ [12pt]
      & & + af_\pi^2 {\rm tr} \left[V_\mu-\frac{1}{2i} ( {\cal D}_\mu \xi_L
      \cdot \xi_L^\dagger + {\cal D}_\mu \xi_R \cdot \xi_R^\dagger)  \right]^2
      + {\cal L}_{\rm kin} ( V_\mu ,{\cal B}_\mu),
\end{eqnarray}
where the pion decay constant $f_\pi=93$ MeV.
The covariant derivative, ${\cal D}_\mu \xi_{L,R}$, is given by
\begin{eqnarray}
{\cal D}_\mu \xi_{L,R}= \partial_\mu \xi_{L,R}+ ie \xi_{L,R} {\cal B}_\mu
 \tau_3/2.
\end{eqnarray}
In the ``unitary" gauge,
\begin{eqnarray}
\xi_L^\dagger(x)=\xi_R(x)=e^{i\pi(x)/f_\pi} \equiv \xi(x),
\end{eqnarray}
and after rescaling $V_\mu \rightarrow gV_\mu$,
the effective lagrangian takes the form,
\begin{eqnarray}\label{lag}
{\cal L} & = &\frac{1}{4} {\rm tr}(\partial_\mu U \partial^\mu U^\dagger)
     -\frac{1}{4}(\partial_\mu {\cal B}_\nu- \partial_\nu {\cal B}_\mu)^2
     -\frac{1}{4}(\partial_\mu V_\nu- \partial_\nu V_\mu)^2
     +\frac{1}{2} m_\rho^2 V_\mu^2\nonumber \\ [12pt]
  & &+g_{\rho \pi \pi} V^\mu \cdot (\pi \times \partial_\mu \pi)
     -eg_\rho V_3^\mu {\cal B}_\mu
     +g_{\gamma \pi \pi}   {\cal B}^\mu  (\pi \times \partial_\mu \pi)_3
      +{\cal L}_{>3},
\end{eqnarray}
where $U=\xi_L^\dagger\xi_R=\xi^2(x)$ and ${\cal L}_{>3}$ are
high order terms involving
more than three fields.  The parameters in eq. (\ref{lag})
are given as
\begin{eqnarray}
m_\rho^2 &=& ag^2 f_\pi^2,\\
g_\rho &=& a g f_\pi^2,\\
g_{\rho \pi \pi} &=& \frac{1}{2}a g,\\
g_{\gamma \pi \pi} &=& (1-\frac{1}{2}a)e.
\end{eqnarray}
For $a=2$, these formula are known to
give automatically the universality of $\rho$-couplings ($g_{\rho\pi\pi}=g$),
the KSRF relations
\begin{eqnarray}
g_\rho &=& 2 f_\pi^2 g_{\rho \pi \pi}\quad :\quad {\rm KSRF \> (I)},\\
m_\rho^2 &=& 2 g_{\rho \pi \pi}^2 f_\pi^2 \quad : \quad {\rm KSRF \> (II),}
\end{eqnarray}
and the $\rho$ meson dominance of the
pion electromagnetic form factor ($g_{\gamma \pi \pi}=0$).


In this effective lagrangian,
the pion electromagnetic form factor in free space can be obtained at tree
level from the diagram shown
in Fig. 1.   One sees that the vector meson dominance appears naturally
so a photon converts into a vector meson which then interacts with
the pion.  The actual process of probing the pion
is thus related more to the propagation of
the vector meson than to the intrinsic
size of the pion itself. The resulting pion electromagnetic form factor
is exactly the same as eq. (1) assumed in VMD.
The temperature effect on the pion electromagnetic form factor is
obtained by taking thermal loops into consideration.
We include only one-loop diagrams as
the hadronic matter is rather dilute at temperatures considered here.
For simplicity, we carry out the calculation in the rest
frame of the photon and neglect the small finite pion mass.

First, we consider the temperature effect on the photon-vector-meson
coupling as shown in Fig. 2a.
This correction is related to
the mixing of the vector and axial vector current correlators
in hot matter.
Using the relation
\begin{equation}
\int^\infty_0 {dx\over\pi^2}{x\over e^{x/T}-1}={T^2\over 6},
\end{equation}
we have
\begin{equation}
F_\pi^{(1)}(q,T)=-F_\pi(q^2){\epsilon\over2},
\end{equation}
where $q$ is the photon four momentum,
$\epsilon={T^2/ 6f_\pi^2}$,
and $F_\pi(q^2)$ is the
pion electromagnetic form factor in free space given by eq.~(1).
We see that, contrary to the modification of the
vector meson mass in hot matter,
the pion electromagnetic form factor has a $T^2$ dependence correction.

The change of rho meson properties
in hot matter gives the correction in Fig. 2b, i.e.,
\begin{equation}
F_\pi^{(2)}(q,T)=-F_\pi(q^2){m_\rho^2\over q^2-m_\rho^2}G_1(q,T),
\end{equation}
where
\begin{equation}
G_1(q,T)={1\over6f_\pi^2}\int^\infty_0{dx\over\pi^2}
{1\over e^{x/T}-1}{x^3\over x^2-q^2/4}.
\end{equation}

In Figs. 2c-2f, we show the vertex corrections.
The contribution from Fig. 2f is suppressed by the
large vector meson mass and
will be neglected. The other contributions can be written as
\begin{equation}
F_\pi^{(3)}(q,T)=F_\pi(q^2)[F_\pi^{(c)}+F_\pi^{(d)}+F_\pi^{(e)}],
\end{equation}
where
\begin{eqnarray}
F_\pi^{(c)}(q,T)&=&{m_\rho^2\over 2f_\pi^2q^2}[G_2(q,T)-G_3(q,T)],\\
F_\pi^{(d)}(q,T)&=&-G_1(q,T),\\
F_\pi^{(e)}(q,T)&=&-{5\over 24}\epsilon.
\end{eqnarray}
In the above, $G_2(q,T)$ and $G_3(q,T)$ are given by
\begin{eqnarray}
G_2&=&{1\over4\pi^2}\int^\infty_0 dx{1\over e^{x/T}-1}
\left[4x-{m_\rho^2\over q}\ln\left({m_\rho^2+2qx\over m_\rho^2-2qx}\right)
+x\ln\left({m_\rho^4\over m_\rho^4-4q^2x^2}\right)\right],\\
G_3&=&{1\over4\pi^2}\int^\infty_0 dx{x\over e^{x/T}-1}(K_++K_-),
\end{eqnarray}
with
\begin{equation}
K_\pm={1\over q({q\over2}\mp x)}({3\over2}q^2+m_\rho^2\pm qx)
\left[2\pm{m_\rho^2\pm qx\over qx}
\ln\left({m_\rho^2\over m_\rho^2\pm 2qx}\right)\right].
\end{equation}
Contributions from thermal vector mesons are very small
due to their large masses in the Boltzmann factor and are neglected in
the study.

In a medium, a pion changes its properties through
interactions with thermal pions and vector mesons
as shown in Fig. 2g and 2h. This modification of the form factor due to
the pion wave function renormalization at finite temperature is given by
\begin{equation}
F_\pi^{(4)}(q,T)=F_\pi(q^2)Z_\pi,
\end{equation}
with
\begin{equation}
Z_\pi=-{\epsilon\over6}+{m_\rho^2\over 8f_\pi^2}{1\over q}
        \int^\infty_0{dx\over\pi^2}{1\over e^{x/T}-1}
        \ln\left({m_\rho^2+2qx\over m_\rho^2-2qx}\right).
\end{equation}

In the hidden gauge approach, there is also
direct photon-pion coupling at
finite temperature as shown in Fig. 2i and 2j,
which explicitly modifies the notion of vector meson dominance.
This contribution is given by
\begin{equation}
F_\pi^{(5)}(q,T)={5\over8}\epsilon+{3m_\rho^2\over 2f_\pi^2}G_4(q,T),
\label{f5}
\end{equation}
where
\begin{equation}
G_4(q,T)={1\over 4q^2}\int^\infty_0{dx\over \pi^2}{x\over e^{x/T}-1}
\left[4-{(q^2+2m_\rho^2)\over 2qx}
 \ln\left({m_\rho^2+2qx\over m_\rho^2-2qx}\right)
+\ln\left({m_\rho^4\over m_\rho^4-4q^2x^2}\right)\right].
\end{equation}

The pion electromagnetic form factor at finite temperature is
obtained by adding all contributions shown in Fig. 2.
Fig. 3 shows the $q$ dependence of the pion form factor around the
$\rho$ resonance for different temperatures.
The form factor is seen to be reduced
near the resonance as temperature increases.
We obtain a reduction of the form factor by 40\% at $q\sim m_\rho$
when $T=180$ MeV. This result is comparable with that obtained
using the QCD sum-rule approach which shows that
at $q^2\sim (1 {\rm GeV})^2$
the form factor at $T\sim$ 0.9 $T_c$ is about half its value at
$T=0$ \cite{domi}.
It is also consistent with that based on the perturbative QCD at high $q^2$
\cite{satz}.

The reduction of the pion electromagnetic form factor
at finite temperature
is mainly due to the modification
of the photon-vector meson
coupling, the increase of the $\rho$-meson width,
and the vertex corrections.
These are probably related to chiral symmetry restoration and
the deconfinement phase transition in hot hadronic matter.
The photon-$\rho$-meson coupling is modified due to the isospin
mixing at finite temperature which has been regarded as a possible
signature for the
partial restoration of chiral symmetry in hot matter \cite{ks}.
The resonance width has also been expected to increase in hot
hadronic matter as the system undergoes chiral symmetry restoration
and the deconfinement phase transition \cite{pisaski}.
Since
\begin{equation}
\Gamma(\rho\to 2\pi)\sim {g^2_{\rho\pi\pi}p_\pi^3\over m_\rho^2}
                    \sim {p_\pi^3\over f_\pi^2},
\end{equation}
the $\rho$-meson width will increase as $f_\pi\to 0$ when the temperature
is close to the critical value.
The vertex corrections, which lead to a reduction of the
rho-pion coupling constant at finite temperature,
may be related to the recent suggestion that the
pion-vector meson coupling constant vanishes when chiral symmetry
is restored in the vector limit \cite{rho}.
The possible relation between the suppression of the form factor
and phase transition in hot hadronic matter also has been suggested via
QCD sum rules \cite{domi} and the QCD factorization formula \cite{satz}
to lead to similar suppressions in the pion electromagnetic form factor.


We have calculated the dilepton production rate from pion-pion annihilation
in hot hadronic matter.
The production rate of dileptons with vanishing three momentum
in hot hadronic matter is then given by \cite{gale}
\begin{equation}
{d^4R\over d^3qdM}\Biggr\vert_{\vec{q}=0}={\alpha^2\over3(2\pi)^4}
{\vert F_\pi(M,T)\vert^2\over(e^{\omega/T}-1)^2}
\sum_k{k^4\over\omega^4}\Biggl\vert{d\omega\over dk}\Biggr\vert^{-1},
\label{dilep}
\end{equation}
where $M$ is the dilepton invariant mass.
The momentum and energy of the pion are denoted by $k$ and $\omega$,
respectively, and are related by its dispersion relation in the medium.
The last factor takes into account this effect.
The sum over $k$'s is restricted by $\omega(k)=M/2$.
However, the modification of the pion dispersion relation at finite
temperature is small \cite{song2} and will be neglected.

The dilepton production rate
is shown in Fig. 4 for $T=$ 180.
The result obtained with the modified pion form factor (solid line) is
compared with that calculated using the form factor in free space (dotted
line).
Since the production rate is proportional to the
square of the form factor, we obtain a larger reduction
with temperature in the dilepton
production rate than in the form factor above.
Near the $\rho$ meson resonance we have $dR[F_\pi(M,T)]$ $\sim$
$(3/5)^2$ $dR[F_\pi(M,0)]$ at $T=180$ MeV,
and the dilepton production rate is
reduced by almost a factor of three.


Recent experiments at CERN have reported
the ``unaccounted" excess of dileptons in central S-W nuclear collisions,
compared to pA and peripheral collisions \cite{dilep-e}.
The data might be described by assuming,
as for photon production \cite{sinha}, that
a thermalized quark-gluon plasma is formed in the collision,
which then cools due to expansion and hadronizes before
undergoing a freeze-out.
In this model, the most important contribution to dilepton production
comes from the hadronic component of the mixed phase
at $T_c=160\sim 180$ MeV \cite{sinha,sx}.
Our results, however, imply a suppression in the production rate
due to the temperature dependence of the form factor.
This may suggest that other mechanisms are contributing to
photon and dilepton production from hot hadronic matter.
One possibility is the
slow expansion model \cite{sx} in which a mixed phase
expands very slowly and produces more dileptons to compensate for the
suppression due to the modification of the form factor \cite{sem}.

Also, it is of interest to extend present calculations to
the $SU(3)$ limit and
to study the temperature dependence of the form factor
needed in $\bar K-K$ annihilation. This will be relevant to
the double phi meson peak in the
dilepton spectrum, which has recently been suggested as a possible
signal for the phase transition in hot matter \cite{yuki}.
The second phi peak in the dilepton spectrum is from the decay of phi mesons
in the mixed phase, which have reduced masses as a result of
partial restoration of chiral symmetry.

In summary, we have studied the pion electromagnetic
form factor in hot hadronic matter
using an effective lagrangian with vector mesons.
We find that there is a reduction in the magnitude of the form
factor, which could be
understood in terms of the partial restoration of chiral symmetry and
the deconfinement transition in hot hadronic matter.  The reduction in the
electromagnetic form factor
leads to a suppression of dilepton production from
pion-pion annihilation in hot matter.  This effect needs to be included in
future studies of dilepton production from heavy ion collisions.

\bigskip

The work of C.S. and C.M.K. was supported in part
by the National Science Foundation
under Grant No. PHY-9212209 and the Robert A. Welch Foundation under
Grant No. A-1110.
The work of S.H.L. was supported by the Basic Science Research
Institute Program, Korea Ministry of Education, BRSI-94-2425,
KOSEF through CTP in Seoul National University and Yonsei
University research grant.

\newpage

\centerline{{\bf Figure Captions}}

\vspace{1.5cm}

\noindent
{\bf Fig.\ 1:} The pion electromagnetic form factor at tree level.
\vspace{1cm}

\noindent
{\bf Fig.\ 2:} One-loop corrections to the pion electromagnetic form factor.
 Solid, wavy, and dotted lines denote, respectively,
 the vector meson, the photon and the pion.
\vspace{1cm}

\noindent
{\bf Fig.\ 3:} The pion electromagnetic form factor at finite temperature.
\vspace{1cm}

\noindent
{\bf Fig.\ 4:} The thermal dilepton production rate from
pion-pion annihilation at
$T$= 180 MeV. Solid and dotted lines are results obtained with modified
and free form factors, respectively.


\begin{thebibliography}{99}
\bibitem{sem} In ref.~\cite{sx} the slow expansion model
              has been suggested to explain the experimental
              data for photon and dilepton production,
              which has been shown to be well described
              by the conventional model for photon production \cite{sinha},
              without including the suppression of the form factor at
              finite temperature.
\end{thebibliography}
\end{document}